\documentclass[a4paper,11pt,preprint,twocolumn,5p,sort&compress]{elsarticle}
\makeatletter
\def\ps@pprintTitle{%
 \let\@oddhead\@empty
 \let\@evenhead\@empty
 \def\@oddfoot{\hfil \thepage \hfil}%
 \let\@evenfoot\@oddfoot}
\makeatother

\usepackage{amssymb,latexsym,amsmath}

\usepackage{mathrsfs}

\begin{document}

\begin{frontmatter}

\title{Chromo-Natural inflation in Supergravity}

\author{Gianguido Dall'Agata}

\address{Dipartimento di Fisica e Astronomia ``Galileo Galilei''\\
 		Universit\`a di Padova, Via Marzolo 8, 35131 Padova, Italy}

\address{INFN, Sezione di Padova \\
		Via Marzolo 8, 35131 Padova, Italy}

\begin{abstract}
	We present supergravity realizations of chromo-natural inflationary models.
	We show that by using superpotentials with ``imaginary'' holomorphic functions of the inflaton one can obtain effective theories of inflation that are also consistent truncations of the original supergravity models.
\end{abstract}


\end{frontmatter}

\renewcommand{\arraystretch}{1.2}



\section{Introduction} 
\label{sec:introduction}

Natural inflationary models have the nice property of protecting the required flatness of the inflationary potential against radiative corrections by means of an approximate shift symmetry \cite{Freese:1990rb}.
For instance, in the case of an axion field $\phi$ acquiring a potential from instanton corrections, the symmetry can be broken in a controlled way:
\begin{equation}
	\label{axionpotential}
	V(\phi) = \Lambda^4 \left(1- \cos\frac{\phi}{f_\phi}\right).
\end{equation}
While these models have been studied for many years (see \cite{Pajer} for a review), only recently they have been generalized to Supergravity, keeping the stabilization of the other fields under control \cite{Kallosh:2014vja}.
All these models, however, suffer the fact that a greater than planckian axion decay constant $f_{\phi}$ is needed, which seems to be very difficult to embed in string theory \cite{Banks:2003sx,Baumann:2014nda}.
Modifications of this scenario to solve this problem have been proposed \cite{Kim:2004rp}, usually by introducing additional axion fields or by creating conditions for a transplanckian evolution of the inflation or by slowing down the axion evolution through particle production.
This last mechanism has been used in inflationary scenarios involving gauge fields \cite{Maleknejad:2011jw} and in particular in the so-called chromo-natural inflation scenario \cite{Adshead:2012kp}, where the axion couples to the topological term of an SU(2) triplet of gauge fields:
\begin{align}
	\label{chromoaction}
	S =& \int d^4 x \sqrt{-g}\left[\frac{M_P}{2}R - \frac12 \partial_\mu \phi \partial^\mu \phi -V(\phi) \right.\\[2mm]
	&\left.-\frac14\,{\rm Tr}(F_{\mu\nu}F^{\mu\nu})+ \frac{g_{\phi}}{8 \sqrt{-g}}\, \frac{\phi}{f_\phi} \, \epsilon^{\mu\nu\rho\sigma}{\rm Tr}(F_{\mu\nu} F_{\rho \sigma})\right]. \nonumber
\end{align}
If the vector fields acquire an isotropic vacuum expectation value during inflation\footnote{Here and in what follows $I,J=1,2,3$ are the SU(2) gauge indices of the adjoint representation and $a=1,2,3$ labels the space coordinates $x^\mu = \{x^0, x^a\}$.}
\begin{equation}
	A_0^I = 0, \quad A^I_a = \delta^I_a\, Q(t),
\end{equation}
the modification to the axion equations of motion are such that the inflaton can be in slow roll also for potentials that would otherwise be too steep to support inflation.
In addition to the force from the axion potential there is in fact a magnetic force proportional to $g_{\phi}$, acting as a friction term and hence granting the slow roll of the inflaton.
While this scenario allows for sub-planckian $f_{\phi}$, the number of e-folding that can be obtained depends crucially on $g_{\phi}$, which should be sufficiently large \cite{Baumann:2014nda}.

Also this model has been studied for various years and various modifications to make it compatible with the observed data have been proposed (see for instance  \cite{Adshead:2016omu}).
However, a Supergravity embedding, which is a first step to find its embedding in an ultraviolet complete theory of quantum gravity like string theory, is still missing.
In this note we fill this gap, by providing Supergravity models that consistently reproduce various models of chromo-natural inflation as effective theories.
We also comment on the possible generation of the various ingredients from string theory.


\section{Natural inflation in Supergravity} 
\label{sec:natural_inflation_in_Supergravity}

The first step in our construction is the embedding of natural inflationary models in Supergravity.
The first example of such embedding was provided in \cite{Kallosh:2014vja}, where various Supergravity models have been constructed, leading to effective theories compatible with natural inflation.
In all the models presented in \cite{Kallosh:2014vja} the superpotential is linear in the goldstino superfield $S$ and the K\"ahler potential is a function of only the real or imaginary part of the inflaton superfield $\Phi$, so as to avoid the Supergravity $\eta$-problem.
While most of these models do not give consistent truncations to the inflaton alone, the sgoldstino acts as a stabilizer, generating a large effective mass for the partner of the axion and for itself, so that one can produce effective models with an axion potential of the form (\ref{axionpotential}).

In what follows we first review and refine the construction presented in \cite{Kallosh:2014vja} by building Supergravity models that allow for effective theories of natural inflation that are also consistent truncations, i.e.~such that solutions to the effective theory equations of motion are also exact solutions of the full model.
We do this following the ideas presented in \cite{Kallosh:2010xz}.
Since we will also be interested in models with two light inflaton fields, we introduce two K\"ahler potentials, with the same structure, depending on the inflatons $\Phi_i$ and on the stabilizers $S_i$ as
\begin{equation} \label{Kpot}
	K_i = \frac12\, (\Phi_i+\bar \Phi_i)^2 + S_i \bar S_i - \frac{b}{M_P^2} \,(S_i \bar S_i)^2.
\end{equation}
Introducing canonically normalized fields
\begin{equation}
	\Phi_i = \frac{1}{\sqrt2} \, \left(\alpha_i + i \,\phi_i\right),
\end{equation}
we see that the K\"ahler potentials (\ref{Kpot}) and consequently the scalar $\sigma$-models have shift symmetries $\phi_i \to \phi_i + c_i$, while the parameter $b$ is introduced for stabilization of $S$ at $S=0$.
We also introduce a superpotential that is a linear combination of
\begin{equation}\label{supo}
	W_i =  S_i\ {\mathscr F}_i\left(\frac{\Phi_i}{f_\phi}\right),
\end{equation}
where ${\mathscr F}_i$ are ``imaginary" holomorphic functions, namely they satisfy
\begin{equation}
	\overline{{\mathscr F}_i(z)} = {\mathscr F}_i(-\bar z).
\end{equation}
Whenever these functions can be given in terms of a power series, their general expansion is 
\begin{equation}\label{expansion}
	{\mathscr F}_i(z) = \sum_n a_n (i z)^n, 
\end{equation}
for $a_n \in {\mathbb R}$.
The objective is to have a scalar potential that is even in the $\alpha_i$ fields so that we can consistently truncate them to zero.
Given the structure of the K\"ahler potential (\ref{Kpot}) and superpotential (\ref{supo}), the same result can be obtained if all the $a_n$ coefficients are imaginary.
This corresponds to the constraint $\overline{{\mathscr F}_i(z)} = -{\mathscr F}_i(-\bar z)$ and the extra imaginary factor in ${\mathscr F}_i$ can be safely reabsorbed in the corresponding $S_i$ field, without changing the resulting model.
Altogether these ingredients give a scalar potential
\begin{equation}
	V = e^{K/M_p^2} \left(g^{i \bar \jmath}D_iW \, D_{\bar\jmath}\overline{W} - 3 \frac{|W|^2}{M_p^2}\right),
\end{equation}
which can be consistently truncated to configurations where $S_i = 0 = \Phi_i + \bar{\Phi}_i$.
In fact the scalar potential depends at least quadratically on the $S_i$ fields and it is even with respect to the $\alpha_i$ fields, i.e.
\begin{equation}
	\partial_{S_i} V|_* = \partial_{\alpha_i} V|_* = 0,
\end{equation}
where $|_*$ denotes the evaluation of the quantity at $S_i = \alpha_i = 0$.
The truncated scalar potential is then
\begin{equation}
	\label{Vtrunc}
	V|_* = \sum_i  \left|{\mathscr F}_i\left(\frac{i}{\sqrt2} \frac{\phi_i}{f_{\phi}}\right)\right|^2
\end{equation}
and the masses of the other fields along the $\phi_i$ directions are
\begin{align}
	\frac{M_p^2\,m_{Re\, S_i}^2}{V|_*} &= \frac{M_p^2\,m_{Im\, S_i}^2}{V|_*} =
	 4b - \frac{M_p^2}{f_{\phi}^2} \left[\frac{{\mathscr F}_i'|_*}{{\mathscr F}_i|_*}\right]^2, \nonumber\\[2mm]
\frac{M_p^2 \, m_{\alpha_i}^2}{V|_*} &= 2 - \frac{M_p^2}{f_{\phi}^2}\left[\frac{{\mathscr F}_i^\prime|_*}{{\mathscr F}_i|_*}\right]^2
	+ \frac{M_p^2}{f_{\phi}^2}\frac{{\mathscr F}_i''|_*}{{\mathscr F}_i|_*}.
\end{align}
These are naturally of the order of the Hubble parameter for choices of ${\mathscr F}_i$ leading to inflationary potentials and for choices of $b$ that do not require fine tuning.

To reproduce the scalar potential (\ref{axionpotential}) we can take only the first copy of $S$ and $\Phi$ fields and set ${\mathscr F}_1|_* =\sqrt2\, \Lambda^2 \sin\left(\frac{\phi_1}{2f_{\phi}}\right)$, which means
\begin{equation}\label{Fchromo1}
	{\mathscr F}_1(\Phi_1) = i\, \sqrt2\,\Lambda^2 \sinh \left(\frac{1}{\sqrt2}\,\frac{\Phi_1}{f_\phi}\right).
\end{equation}
This corresponds to Model 3 in \cite{Kallosh:2014vja}, which provides indeed a consistent truncation and not just an effective theory of natural inflation.
The other options presented in \cite{Kallosh:2014vja} do not lead to consistent truncations, but they give otherwise well-defined effective theories.

More options are available if one uses non-linear representations of supersymmetry.
As shown in recent times, one has more functional freedom when using non-linear supersymmetry in Supergravity and one can therefore accommodate more easily inflationary models, satisfying all consistency requirements \cite{Nilpotent}.
For instance one could obtain a scalar potential of the form (\ref{Vtrunc}) by introducing two constrained chiral superfields $X$ and $Y$, satisfying $X^2 = 0 = XY=0$ in a model with K\"ahler potential \cite{DallAgata:2015pdd}
\begin{equation}\label{kahlnil}
	K = \frac12 (\Phi_1 + \bar \Phi_1)^2 + \frac12 (\Phi_2 + \bar \Phi_2)^2 + X \bar X + Y \bar Y
\end{equation}
and superpotential
\begin{equation}\label{suponil}
	W = \Lambda^2\, \left( X {\mathscr F}_1 + Y {\mathscr F}_2\right).
\end{equation}
Actually, one could also constrain the $\Phi_i$ fields by setting $X (\Phi_i +\bar{\Phi}_i)=0$ and produce a direct truncation to the axion fields $\phi_i$ \cite{Ferrara:2015tyn}, while at the same time removing the constraint on the ${\mathscr F}_i$ functions to be ``imaginary'' holomorphic.
This in turn would allow for an easy embedding as consistent truncations also of the other models presented in \cite{Kallosh:2014vja}.
The only delicate point in doing this is that consistency requires supersymmetry to be broken also at the exit of inflation, which implies that ${\mathscr F}_i \neq 0$ everywhere in field space.
This is however a small adjustment that can be easily incorporated in the description by adding a sufficiently small constant term in (\ref{suponil}).


\section{Chromo-natural inflation in Supergravity} 
\label{sec:chromo_natural_inflation_in_Supergravity}

The construction of Supergravity models of chromo-natural inflation like (\ref{chromoaction}) requires the introduction of three vector multiplets in the theory, gauging an SU(2) group.
This adds a number of terms to the lagrangian, including
\begin{equation}
	-\frac14\, {\rm Re}f_{IJ} F^I_{\mu\nu}F^{J\,\mu\nu} + \frac18\, {\rm Im}f_{IJ}\, \frac{\epsilon^{\mu\nu\rho \sigma}}{\sqrt{-g}} \, F^I_{\mu\nu} F^J_{\rho \sigma},
\end{equation}
where $f_{IJ}$ is a matrix, which is also a holomorphic function of the chiral multiplets.

The simplest model of chromo-natural inflation given in \cite{Adshead:2012kp} can then be embedded by choosing
\begin{equation}\label{gaugekin}
	f_{IJ}(\Phi_1) = \delta_{IJ} \left(1 + \sqrt2\,g_{\phi}\,\frac{\Phi_1}{f_\phi}\right).
\end{equation}
Once we add this term to the K\"ahler potential $K_1$ and superpotential generated by (\ref{Fchromo1}), we obtain that there is a consistent truncation to the axion field such that
\begin{equation}
	{\rm Re}f_{IJ}|_* = \delta_{IJ}, \quad {\rm and} \quad	{\rm Im}f_{IJ}|_* = \delta_{IJ}\, g_{\phi}\, \frac{\phi_1}{f_\phi},
\end{equation}
as expected.
We would like to remind our reader that in Supergravity, the non-trivial imaginary gauge-kinetic function produces along the inflationary trajectory a derivative coupling between the axion and the gaugini $\lambda^I$ and couplings between the vector field-strengths, the gravitino, the gaugini and the axino field $\chi^1$:
\begin{align}
	&\frac{1}{4} \frac{g_{\phi}}{f_\phi}\, \partial_\mu \phi_1\, {\rm Tr}( \bar{\lambda} \gamma_5 \gamma^\mu \lambda ) +\frac{1}{4M_P} {\rm Tr} \left(\bar \psi_\mu \gamma^{\nu \rho}\gamma^\mu \lambda\, F_{\nu \rho}\right)\\[2mm]
	&-\frac14 \frac{g_{\phi}}{f_\phi} \left[{\rm Tr}(F_{\mu\nu} \bar{\chi}_L^1 \gamma^{\mu\nu}\lambda_L) + {\rm Tr}(F_{\mu\nu} \bar{\chi}_R^{\bar{1}} \gamma^{\mu\nu} \lambda_R) \right]. \nonumber
\end{align}
The first term should be of concern for the reheating process, because when the axion picks up speed this coupling gives a new channel of particle production with respect to the standard scenario.

Another model of chromo-natural inflation, proposed in \cite{Obata:2016tmo}, contains two light fields, an axion, with the same couplings as the one of the original model of chromo-natural inflation, and another field, with an exponential scalar potential
\begin{equation}
	V(\phi_2) = \Lambda^4_2 \exp (r\, \phi_2)
\end{equation}
and a non-trivial coupling to the kinetic term of the vector fields via a function $I(\phi_2)^2$.
The embedding of this model in Supergravity requires two chiral fields, with K\"ahler potentials and superpotentials as in the previous section, for
\begin{equation}
	{\mathscr F}_1(\Phi_1) = i\, \sqrt2\,\Lambda^2_1 \sinh \left(\frac{1}{\sqrt2}\frac{\Phi_1}{f_\phi}\right)
\end{equation}
and
\begin{equation}
	{\mathscr F}_2(\Phi_2) = \Lambda^2_2 \exp \left(-\frac{i}{\sqrt2}\, r\, \Phi_2\right).
\end{equation}
In addition, the gauge kinetic function should be
\begin{equation}
	f_{IJ}(\Phi_i) = \delta_{IJ}\left(h(\Phi_2) + \sqrt2\,g_{\phi} \,\frac{\Phi_1}{f_\phi} \right),
\end{equation}
for $h$ another ``imaginary'' holomorphic function, such that $h(\Phi_2)|_* = h\left(\frac{i}{\sqrt2} \phi_2\right) = I(\phi_2)^2$.
Along the truncation we recover the desired couplings
\begin{equation}
	{\rm Re}f_{IJ}|_* = \delta_{IJ} \, I(\phi_2)^2\,,\quad	{\rm Im}f_{IJ}|_* = \delta_{IJ}\, g_{\phi}\, \frac{\phi_1}{f_\phi}.
\end{equation}

As mentioned in the introduction, one of the main problems of the chromo-natural inflationary scenario is the requirement of a fine-tuning of the coupling constant $g_{\phi}$, which should be roughly of ${\cal O}(100)$.
This translates in the requirement of a large coupling in the gauge-kinetic function (\ref{gaugekin}).
A dynamical way to address this is by means of a vev of a second scalar field, in a coupling of the form
\begin{equation}
	f_{IJ}(\Phi_i) = \delta_{IJ} \left(1 + c \,\frac{\Phi_1}{\Phi_2}\right),
\end{equation}
with $c$ a dimensionless constant of order 1.
If the field $\Phi_2$ is stabilized at a value that is a couple of orders smaller than the scale of the axion decay constant one gets the wanted result.
In the setup described before this can be achieved for instance by means of a superpotential function of the form
\begin{equation}
	{\mathscr F}_2\left(\Phi_2\right) = \Lambda_2^2 + \Phi_2^2.
\end{equation}
The field $\Phi_2$ is very rapidly stabilized at $\Phi_2 = i \,\Lambda_2$, with masses during inflation of order the Hubble parameter for all the fields, including the imaginary part of $\Phi_2$ provided $\Lambda_1^2 \sim \Lambda_2 M_P$.
On the inflationary trajectory then we have
\begin{equation}
	{\rm Im}f_{IJ}|_* = \delta_{IJ} \, \frac{\phi_1}{\Lambda_2},
\end{equation}
which implies
\begin{equation}
	g_{\phi} = \frac{f_{\phi}}{\Lambda_2}.
\end{equation}
Overall this gives the scale hierarchy $M_p \gg \Lambda_1 \sim \sqrt{\Lambda_2 M_p} \gtrsim  f_{\phi} \sim 10^2 \,\Lambda_2$, which mildly restricts the allowed parameter space.
It is not too difficult, however, to generalize this simple mechanism to allow for a larger parameter space.
The effect of this mechanism is to create a new vantage point to address the origin and viability of such models because one has to explain the origin of a new perturbative scale $\Lambda_2 \sim 10^{-2} f_{\phi}$ in the superpotential rather than a large value for $g_{\phi}$ in the gauge-kinetic function.

It is interesting to note that also the models of chromo-natural inflation presented in this section can be realized in terms of constrained superfields providing non-linear realizations of supersymmetry.
One simply adds the appropriate gauge kinetic functions presented here to the K\"ahler potential and superpotential (\ref{kahlnil}) and (\ref{suponil}).
One still gets consistent models even in the case where the inflaton superfield is constrained by $X (\Phi -\bar \Phi) = 0$, though in this case the axino terms in the lagrangian disappear in the unitary gauge.

We conclude our presentation with a discussion on the possible string origin of the various ingredients contained in our models.
A rather generic discussion of some possible embedding of chromo-natural inflation within M-theory and string theory has already been presented in \cite{Martinec:2012bv}, where further generalizations of such inflationary scenario have been proposed.
However, in what follows we will assume inflation happens in the regime of validity of effective 4-dimensional Supergravity.
One of the scenarios we have under better control is given by flux compactifications of type II string theory on orientifolds of Calabi--Yau manifolds.
There is a vast body of literature on this, but we will mainly refer to \cite{Grimm:2004uq} and \cite{Jockers:2004yj} for the construction of the Supergravity effective actions, also in the presence of D7-branes.

The first ingredient is the existence of axion-like fields in our effective theory.
This is rather common in string compactifications, due to the existence of a large number of 4-dimensional scalar fields arising from the dualization of the higher-rank form-fields existing in 10 dimensions.
More tricky is the stabilization of the axion partner in the corresponding $N=1$ superfield, but several options are available, like $\alpha'$ corrections to the K\"ahler potentials or flux compactifications generating the appropriate terms in the superpotential, or non-perturbative effects modifying the superpotential, or a combination of all these possibilities.
However, the point of main interest is the generation of the field-dependent terms in the gauge-kinetic functions.
In Calabi--Yau compactifications these terms are a function of the second derivative of the prepotential specifying the Special-K\"ahler geometry underlying the vector multiplet scalar $\sigma$-model.
This relation becomes a direct proportionality in the case of orbifold compactifications \cite{Grimm:2004uq}.
Most Calabi--Yau reductions give origin to a cubic prepotential and therefore it is rather natural to expect gauge kinetic functions depending linearly on some of the scalar fields in the truncation of the $N=2$ vector multiplets.
When also $D7$-branes are present, there are additional couplings to the $D7$-brane gauge degrees of freedom and one gets additional gauge kinetic functions that are linear in the chiral fields related to the volume of the 4-cycles on which such branes are wrapped \cite{Jockers:2004yj}.
Clearly a full consistent uplift requires more work than the simple identification of the possible ingredients, but, as described here, string theory seems to provide rather naturally all the necessary fields and couplings.

In conclusion, we have shown that it is fairly easy to embed models of chromo-natural inflation in $N=1$ Supergravity, because it allows for enough functional freedom to accommodate all the necessary ingredients without many constraints.
A consistent superstring derivation of the presented K\"ahler potentials and superpotentials is more challenging, though we gave explicit directions on their possible origins.
Since the closed string sector can play a crucial role in such determination, a natural development of the analysis presented here is the study of extended Supergravity models, to see if they impose more restrictive constraints on the gauge-kinetic functions.
In fact the analysis of \cite{Kodama:2015iua}, where various inflationary scenarios in maximal Supergravity have been considered, suggests that chromo-natural inflation cannot be embedded at all in such very constrained theories.


\bigskip
\section*{Acknowledgments}

\noindent We would like to thank D.~Cassani, N.~Cribiori, F.~D'Eramo, R.~Kallosh, L.~Merlo and especially M.~Peloso for useful comments and correspondence on this project.
This work is supported in part by the Padova University Project CPDA144437.


\end{document}